# f-P vs P-f based Grid-forming Control under RoCoF Event Considering Power and Energy Limits

Chu Sun, *Member, IEEE*

***Abstract*- Grid-forming (GFM) converter is deemed as one enabler for high penetration of renewable energy resources in power system. However, as will be pointed out in this letter, the conventional power-to-frequency (P-f) GFM control will face a dilemma in keeping power limit and grid synchronization when the energy resource of the converter reaches the limit. To address this challenge, a f-P and Q-V hybrid control is proposed, which exhibits similar GFM performance, particularly under weak grid condition, but is superior in power-limiting and grid synchronization as demonstrated by comparative studies.**

## I. Introduction

In the past decades, as synchronous generators were gradually replaced by renewable energy resources interfaced power electronic converters, a grid-forming control strategy based on power-to-frequency (P-f) negative feedback has been widely discussed [1-2]. However, few researchers considered the power and energy constraints of the DC-side resources. Tracing back to the earliest literature, such control strategies assumed an ideal DC voltage source [3-4]. In reality, the DC side of most grid converters consists of non-dispatchable renewable energy source or energy storage with limited capacity. When power imbalance or frequency disturbance such as RoCoF event occurs, converters will automatically undergo uncontrollable power response to maintain synchronization, which may lead to power overloading or energy depletion, and further synchronization instability [5-6].

This letter analyzes the difficulty P-f grid-forming control to maintain power limiting and synchronization stability simultaneously, when the converter experiences RoCoF event while the resource reaches the power or energy limit. A hybrid f-P and Q-V control is proposed, which can provide grid-forming function while featuring better power-limiting capability and synchronization stability.

## II. Conventional P-f based Grid-forming Control

### A. Coupling of Synchronization and Power Regulation

As illustrated in Fig. 1, for conventional P-f control, active power regulation, grid-synchronization, inertial and damping support, are realized by the same block. When there is grid frequency or phase disturbance, the phase and frequency will vary to catch such change and thus achieve synchronization, but the power will be varied simultaneously due to the coupling relation between frequency and power, as shown in Fig. 2. If a power-limiting block such as saturation is added in the P-f control path, the synchronization process will be blocked and get lost. The parameters of virtual inertia ($H$) and damping ($D$) will affect power-tracking and grid-supporting which are two conflicting objectives, i.e. tight power control needs small $H$ and $D$ which in turn leads to weaker frequency-supporting. In the control, a circular current limiting block may be also added behind the virtual admittance block [7]. In Fig. 2, the small-signal model of power and angle relation is denoted by (1), where $X_g$ is grid impedance, $X_L$ is the equivalent impedance of the converter, and $U_0$ is phase voltage magnitude.

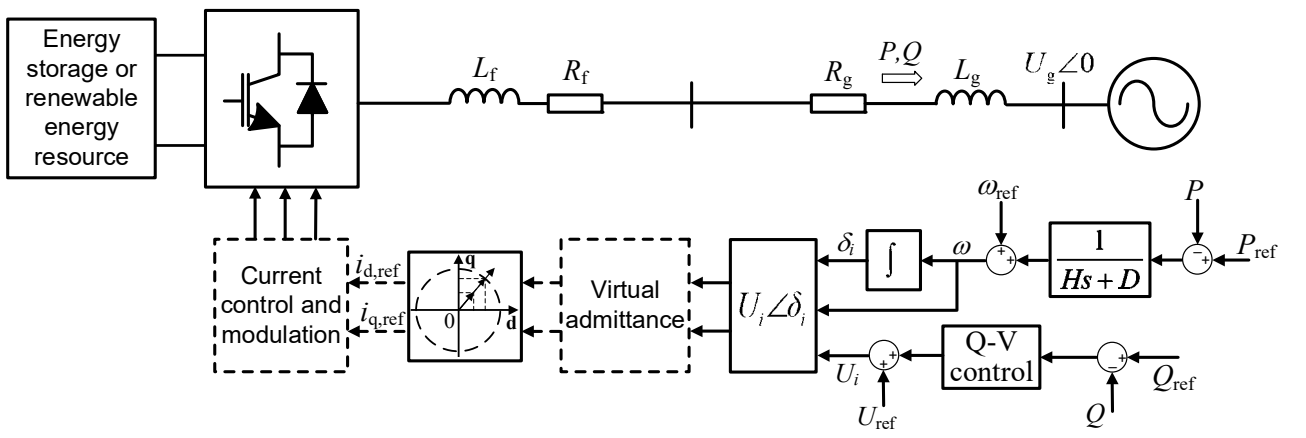

Fig. 1. Grid-connected converter with P-f and Q-V grid-forming control.

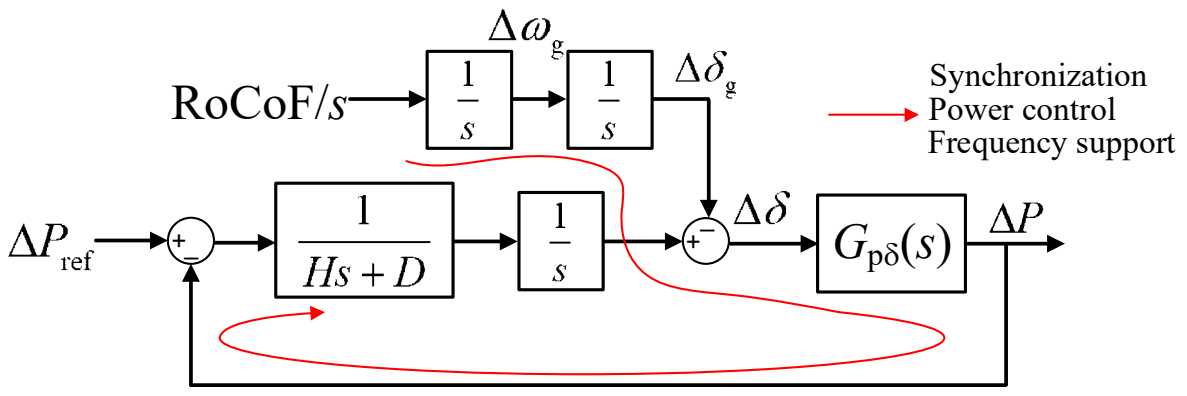

Fig. 2. Simplified control diagram of P-f GFM control under RoCoF event.

$$G_{p\delta} \approx \frac{1.5U_0^2}{X_g + X_L}\cos\delta \tag{1}$$

### B. Difficulty in Power Limiting under RoCoF Event

Due to the ramping change of grid frequency under RoCoF event, for P-f control, it is difficult to maintain power or energy near limit while keeping synchronized, even when P-f control is changed to PI control. In the existing literature, smaller $H$, $D$ or tighter PI control are adopted to limit power when the power or energy comes near the limit. However, the following difficulties must be taken into account.

(1) The control structure and parameters should be adjusted and retuned very carefully to avoid stability issue while ensuring desirable power limiting performance [2].

(2) The control should be adjusted immediately at the right instant, to avoid over-loading or SoC violation. A proper trigger mechanism should be designed to avoid too aggressive or conservative power limiting action [6].

(3) The adjustment should consider the direction of frequency or power variation to avoid mis-operation. For instance, when the SoC or power reaches high limit, downward power response is permissible. However, ordinary P-f PI control cannot guarantee uni-directional power limiting.

(4) Under condition of fluctuating grid frequency while the energy or power of the energy resource comes near limit, there may be frequent switching between normal mode and power-limiting mode. Some hold function should be added to avoid consistent switching between diffierent modes, which leads to more difficulty.

Therefore, implementing P-f control on most converters interfacing renewable energy resource or energy storage with limited energy or power headrooom will be very challenging.

## III. A Hybrid f-P and Q-V Control Scheme

As illustrated in Fig. 3, the control strategy is composed of phase (angle)-shift control with f-P grid supporting control added to the active power reference, and voltage reference is generated by Q-V grid-supporting control [8]. A PLL is used to measure the angle and frequency at PCC for synchronization.

### A. *Decoupling of Synchronization and Power Control*

Assuming the voltage magnitude is near 1 p.u. as close to the grid voltage, the modeling of active power and angle is shown in Fig. 4. Under normal condition, the block $G_{\delta X}$ reflects the linear relation of angle distribution between $X_L$ and grid impedance $X_g$, as expressed by (2) and (3).

As observed, the power control block $G_P$ (namely PI), inertia/damping support block $G_F$ (namely $Hs+D$), and the synchronization block (PLL) are independent and decoupled blocks. Since PLL usually has much faster dynamics compared with $G_F$ and $G_P$, it can be treated as unity ($G_{PLL} \approx 1$) for simplicity. From control disturbance perspective, PLL behaves as a feedforward of the grid frequency/phase disturbance whose variation is therefore counteracted. Even without $G_F$ and $G_P$, the converter can still get synchronized with the grid, with active power being zero.

Before adding f-P control, active power can be tightly regulated at $P_{ref}$ with PI control, and the converter is almost inertia-less and damping-less. The f-P control is a feedforward of the grid disturbance to the power reference, and is cascaded with power control, whose value can be naturally limited by a saturation block, without incurring loss of synchronization.

$$U_p e^{j\delta_{pcc}} \approx G_{\delta X}\left(U_0 e^{j\delta_i} - U_0 e^{j\delta_g}\right) + U_0 e^{j\delta_g}, G_{\delta X} = \frac{X_g}{X_g + X_L} \quad (2)$$

$$\delta_{pcc} \approx G_{\delta X}\left(\delta_i - \delta_g\right) + \delta_g \quad (3)$$

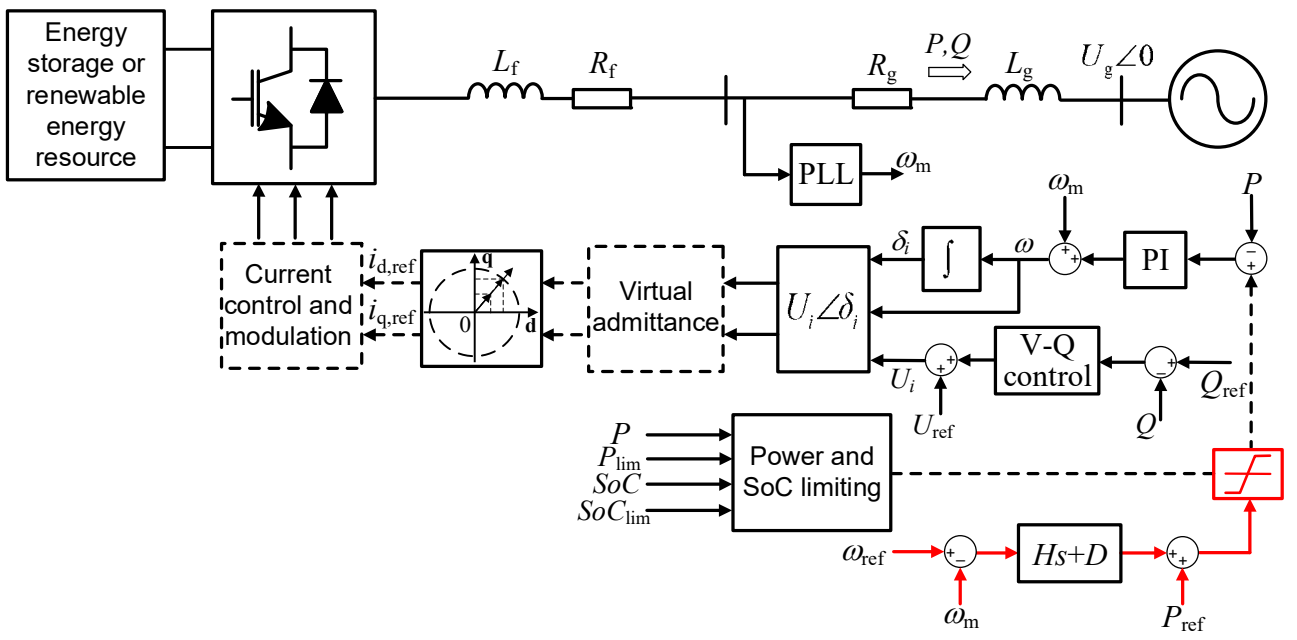

Fig. 3. Converter with f-P control considering energy and power limit.

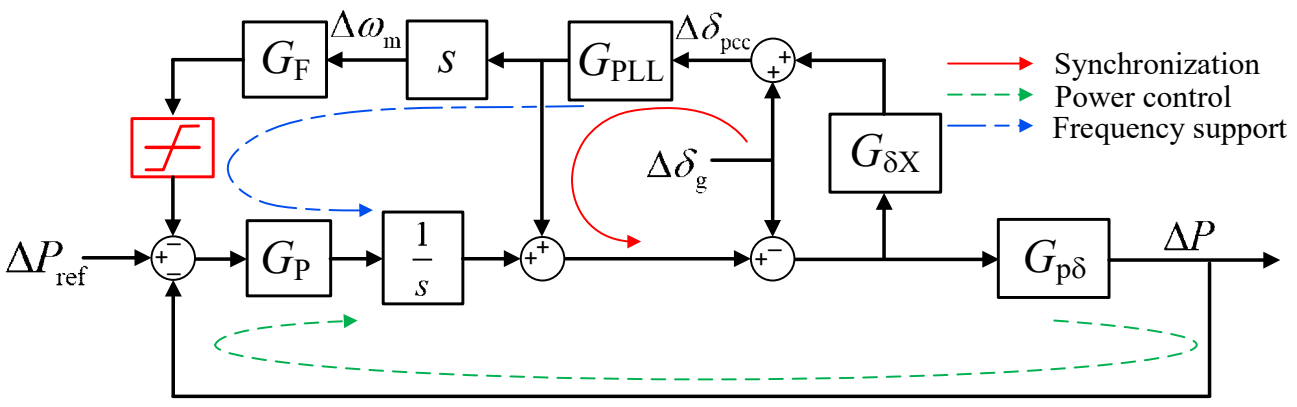

Fig. 4. Simplified active power and synchronization loop of f-P&Q-V control.

### B. *Modeling of Power Control under Weak Grid Condition*

Given large $X_g$ value, $G_{\delta X}$ is close to 1, which means the PCC voltage angle will be mainly determined by the inner voltage angle. Ignoring the saturation block, the control is then approximated as Fig. 5, which is the same as the conventional P-f control diagram in Fig. 2. However, the power and limiting performance will be quite different. The closed-loop transfer function can be derived and omitted here.

### C. *Modeling of Power Control under Strong Grid Condition*

Given small $X_g$ value, $G_{\delta X}$ is close to 0, which means the PLL-measured angle will be mainly determined by the grid-side angle. The simplified control diagram is shown in Fig. 6.

## IV. Control Strategy Comparison and Verification

A system with voltage-source converter interfacing ESS and connected with grid is simulated. A weak grid expecting grid-forming converter installation is considered. The parameters and values of conventional P-f GFM control and the proposed control are listed in Table I. The control performance of ESS near power and SoC limits under RoCoF event is examined.

### A. *Performance of P-f Grid-forming Control*

First, GFM ESS is operating near power limit ($P_{ref}$=0.9 p.u.). A frequency ramping-down (-1 Hz/s to 48 Hz) event is considered. With the current limiting alone to limit power, there will be consistent oscillation in power waveform, indicating loss of synchronization, as observed from Fig. 7.

Near energy limit (high SoC as an example here), the power reference near $P_{ref}$=0 and a frequency ramping-up (+1 Hz/s to 52 Hz) event is considered. Without additional control, ESS will be automatically charged, thus violating the SoC constraint. To avoid this, the current limit behind the virtual admittance is set to a small value near zero (0.012 p.u. here). The result shows that there is consistent power oscillation, indicating loss of synchronization and ineffective power regulation.

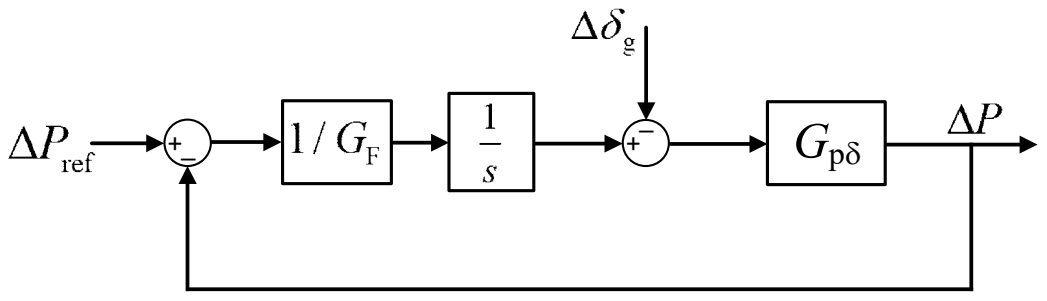

Fig. 5. Simplified active power control diagram under weak grid condition.

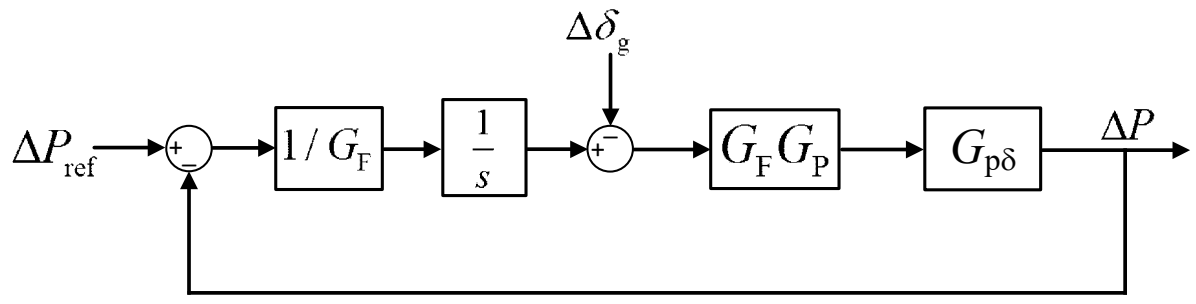

Fig. 6. Simplified active power control diagram under strong grid condition.

Table I. Parameters and Values of the Tested System

| Parameters | Values | Parameters | Values |
|---|---|---|---|
| Base frequency $f_{ref}$ | 50 Hz | Base power | 220 MVA |
| Base voltage $V_{(LL\text{-}rms)}$ | 66 kV | High SoC limit | 0.9 |
| Power rating of ESS | 1 p.u. | Current limit | 1.2 p.u. |
| Energy capacity of ESS | 5 p.u.•s | DC voltage | 185 kV |
| $k_{p,PLL}$ | 800 | $k_{i,PLL}$ | 1500 |
| $R_L$ | 0.01 p.u. | $R_g$ | 0.1 p.u. |
| $X_L$ | 0.05 p.u. | $SCR$ ($1/X_g$) | 1.2 |
| $L_f$ | 0.12 p.u. | $R_f$ | 0.012 p.u. |
| $H$ | 5 s | $D$ | 10 p.u. |
| $k_{p,p}$ | 1 p.u. | $k_{i,p}$ | 5 p.u. |
| $k_{p,i}$ | 0.5 p.u. | $k_{i,i}$ | 16 p.u. |

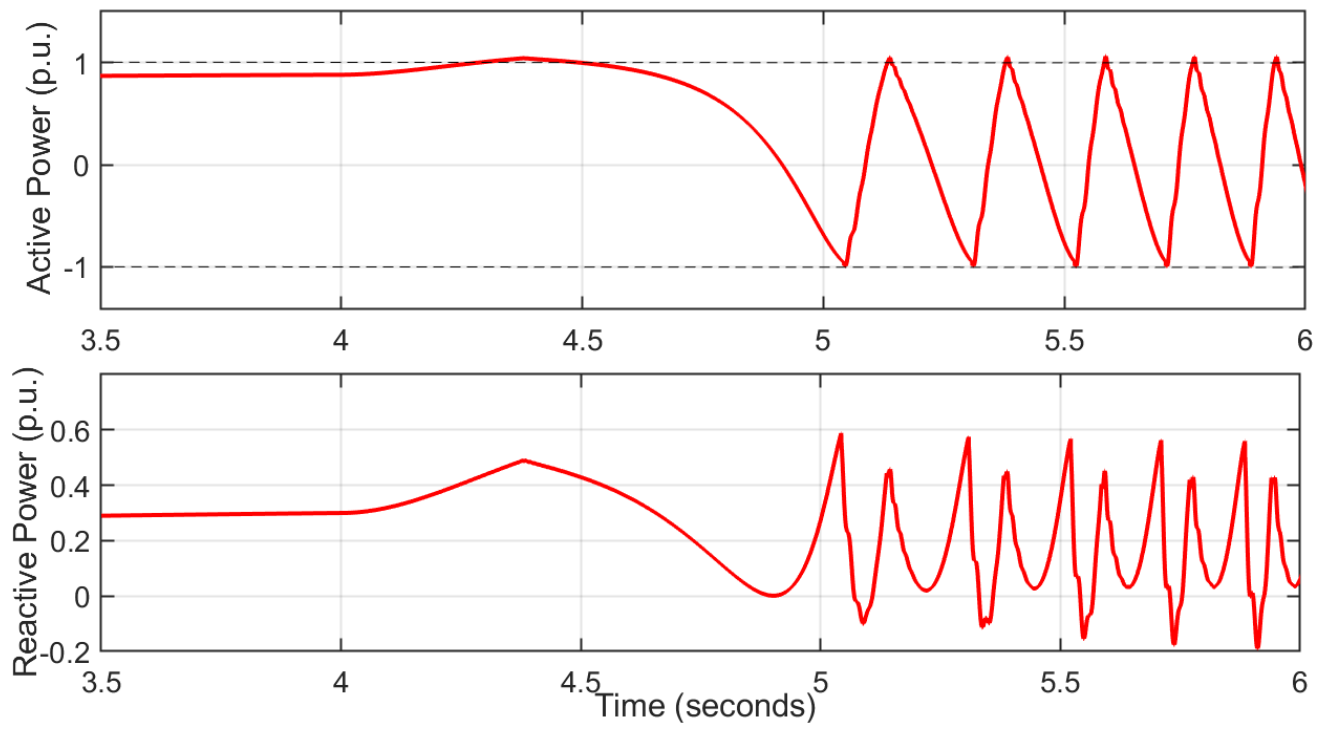

Fig. 7. Power of P-f control for $P_{ref}$=0.9 p.u. under -1 Hz/s RoCoF event.

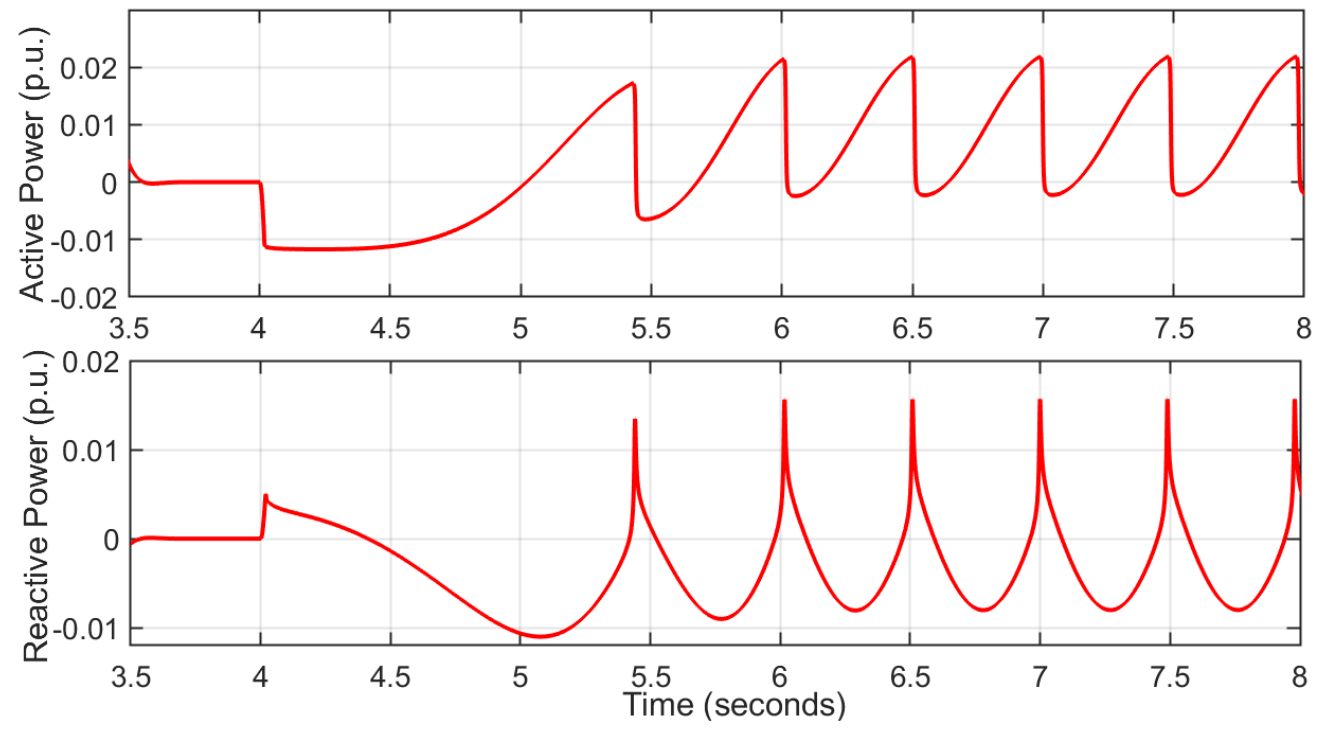

(a)

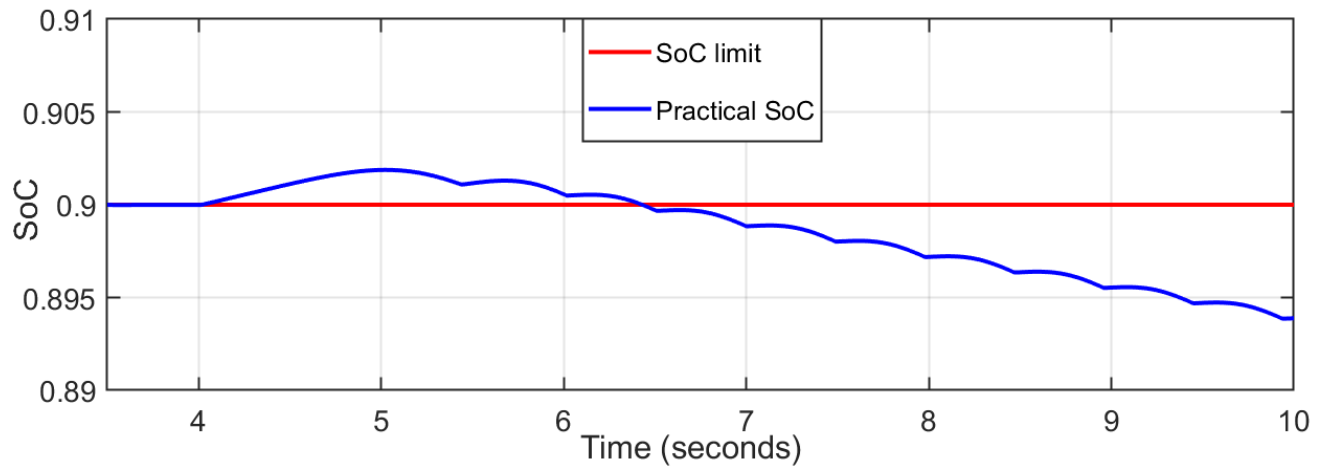

(b)

Fig. 8. Results of P-f grid-forming control for SoC=0.9 under +1 Hz/s RoCoF event. (a) Power waveform. (b) SoC waveform.

## *B. Performance of f-P&Q-V Hybrid Control*

First, the power reference $P_{ref}$=0.5 p.u. and a frequency ramping-down (-1 Hz/s to 48 Hz) event is considered. The active power waveform is shown in Fig. 9, which closely tracks the active power reference generated by f-P control. The equivalent form of P-f control in Fig. 5 gives close result, verifying the effectiveness of the approximate model.

Under various RoCoF events, $P_{ref}$=0.9 p.u. with power limit 1 p.u., and $P_{ref}$=0 p.u. with power limit 0 p.u. is considered. The active power waveforms are presented in Fig. 10(a) and 10(b), respectively, indicating effective power limiting.

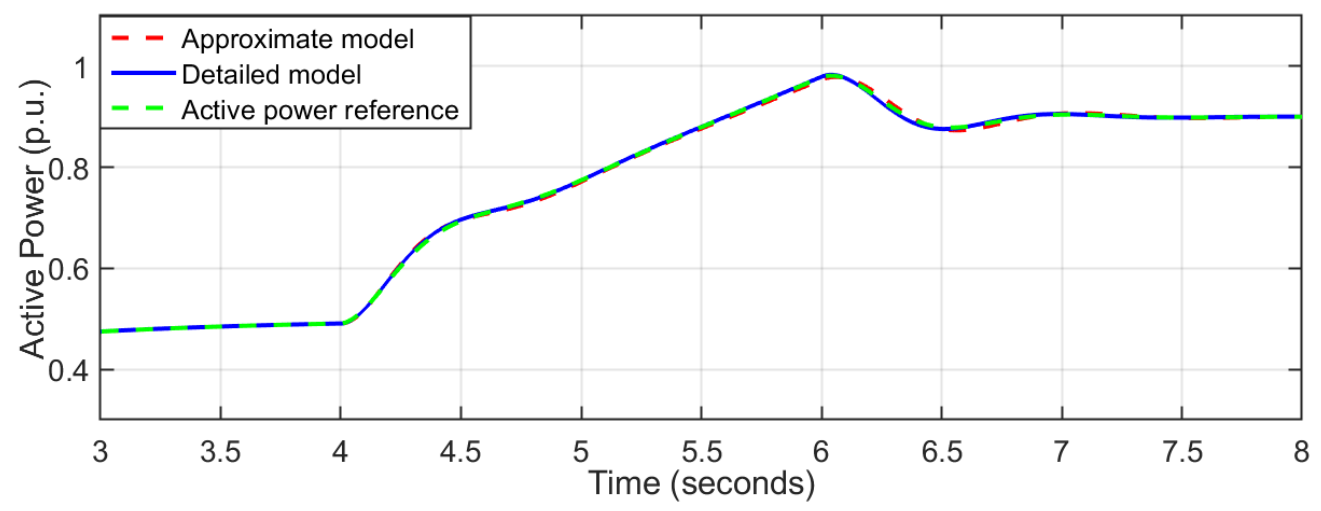

Fig. 9. Power for f-P control with $P_{ref}$=0.5 p.u. under -1 Hz/s RoCoF event.

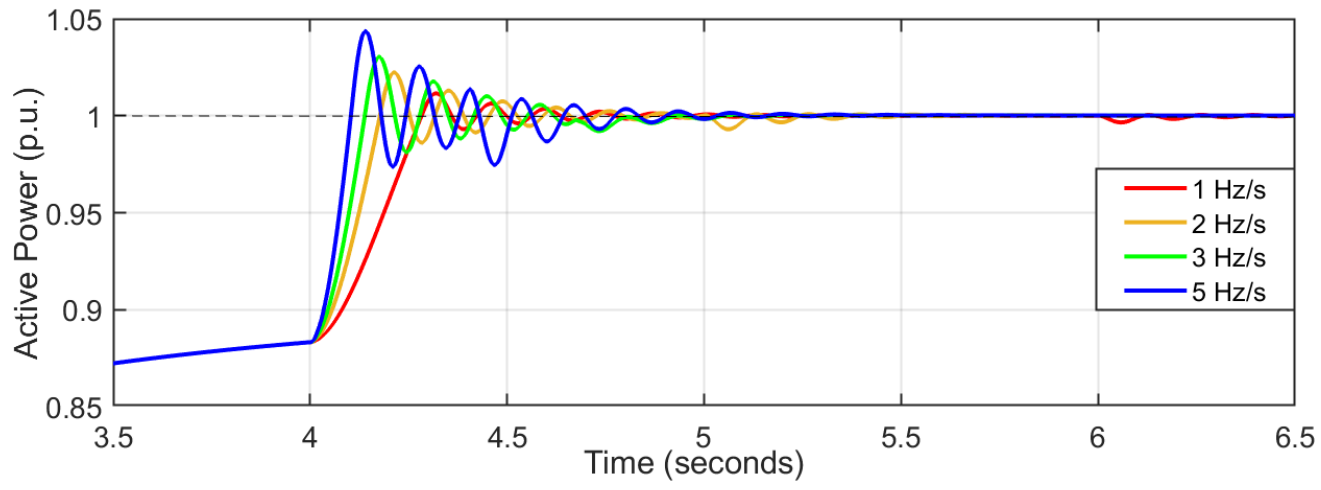

(a)

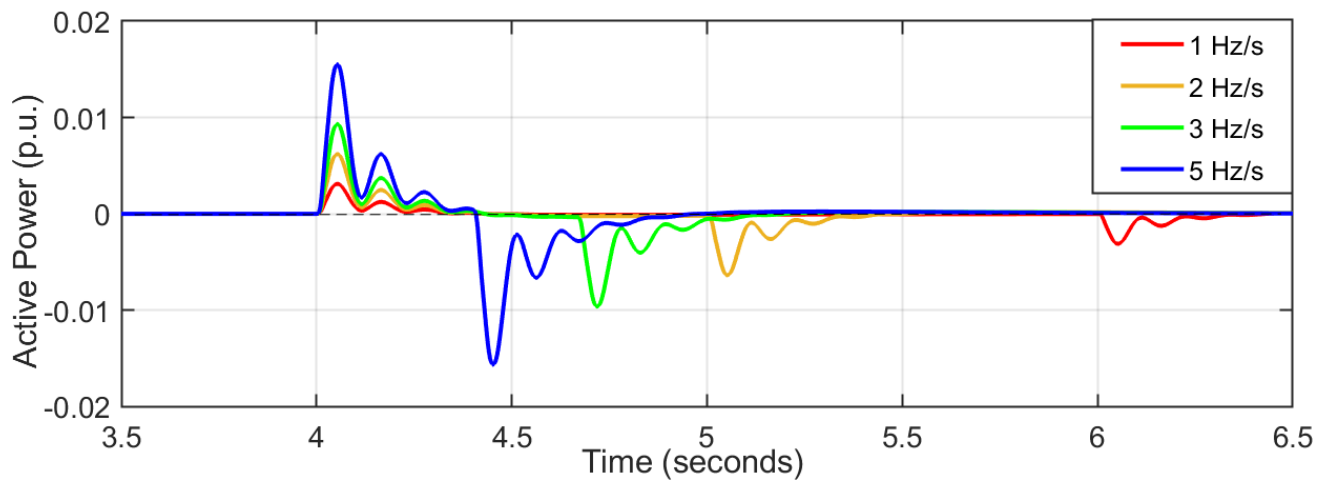

(b)

Fig. 10. Active power for f-P control under various RoCoF events. (a) Power limit at1 p.u. (b) power limit at 0 p.u.

# V. Conclusion

In this letter, the dilemma of power limiting and synchronization of converter with P-f GFM control near power or energy limit is investigated. A workaround, with f-P control hybridized by Q-V control, is proposed. It is revealed the control can be transformed to the conventional “grid-forming” control structure, and can significantly enhance power-limiting capability while keeping grid synchronized under RoCoF event. The proposed control will be more compatible for grid converters interfacing renewable resources and energy storage.

# Reference

[1] J. Rocabert, A. Luna, F. Blaabjerg, P. Rodríguez, "Control of Power Converters in AC Microgrids," IEEE Transactions on Power Electronics, vol. 27, pp. 4734-4749, 2012.

[2] R. H. Lasseter, Z. Chen and D. Pattabiraman, "Grid-Forming Inverters: A Critical Asset for the Power Grid," in IEEE Journal of Emerging and Selected Topics in Power Electronics, vol. 8, no. 2, pp. 925-935, June 2020.

[3] M. C. Chandorkar, D. M. Divan, R. Adapa, "Control of parallel connected inverters in standalone AC supply systems," IEEE Transactions on Industry Applications, vol. 29, pp. 136-143, 1993.

[4] C. Sun, S. Q. Ali, G. Joos and F. Bouffard, "Design of Hybrid-Storage-Based Virtual Synchronous Machine With Energy Recovery Control Considering Energy Consumed in Inertial and Damping Support," in IEEE Transactions on Power Electronics, vol. 37, no. 3, pp. 2648-2666, March 2022.

[5] P. Imgart, A. Narula, M. Bongiorno, M. Beza and J. R. Svensson, "External Inertia Emulation to Facilitate Active-Power Limitation in Grid-Forming Converters," in IEEE Transactions on Industry Applications, vol. 60, no. 6, pp. 9145-9156, Nov.-Dec. 2024.

[6] S. Schneider, R. Morgenstern, K. Lipiec and S. M. Iftekharul Huq, "Inherent phase-based real inertia power response of STATCOM with supercapacitor during high RoCoF events in AC grid," 19th International Conference on AC and DC Power Transmission (ACDC 2023), Glasgow, UK, 2023, pp. 347-351.

[7] B. Fan, T. Liu, F. Zhao, H. Wu and X. Wang, "A Review of Current-Limiting Control of Grid-Forming Inverters Under Symmetrical Disturbances," in IEEE Open Journal of Power Electronics, vol. 3, pp. 955-969, 2022.

[8] G. Joos, L. Moran and P. Ziogas, "Performance analysis of a PWM inverter VAr compensator," in IEEE Transactions on Power Electronics, vol. 6, no. 3, pp. 380-391, July 1991.